\documentclass[sigconf]{acmart}
\AtBeginDocument{%
  \providecommand\BibTeX{{%
    \normalfont B\kern-0.5em{\scshape i\kern-0.25em b}\kern-0.8em\TeX}}}

\begin{document}

\title{Classifying YouTube Comments Based on Sentiment and Type of Sentence}

\author{Rhitabrat Pokharel}
\email{pokharel@pdx.edu}
\affiliation{%
  \institution{Portland State University}
\department{Department of Computer Science}
  \city{Portland}
  \state{OR}
  \country{USA}
}

\author{Dixit Bhatta}
\email{dixit@udel.edu}
\affiliation{%
  \department{Department of Computer and Information Sciences}
  \institution{University of Delaware}
  \city{Newark, DE}
  \country{USA}}

\renewcommand{\shortauthors}{Pokharel and Bhatta}

\begin{abstract}
As a YouTube channel grows, each video can potentially collect enormous amounts of comments that provide direct feedback from the viewers. These comments are a major means of understanding viewer expectations and improving channel engagement. However, the comments only represent a general collection of user opinions about the channel and the content. Many comments are poorly constructed, trivial, and have improper spellings and grammatical errors. As a result, it is a tedious job to identify the comments that best interest the content creators. In this paper, we extract and classify the raw comments into different categories based on both sentiment and sentence types that will help YouTubers find relevant comments for growing their viewership. Existing studies have focused either on sentiment analysis (positive and negative) or classification of sub-types within the same sentence types (e.g., types of questions) on a text corpus. These have limited application on non-traditional text corpus like YouTube comments. We address this challenge of text extraction and classification from YouTube comments using well-known statistical measures and machine learning models. We evaluate each combination of statistical measure and the machine learning model using cross validation and $F_1$ scores. The results show that our approach that incorporates conventional methods performs well on the classification task, validating its potential in assisting content creators increase viewer engagement on their channel.

\end{abstract}




\maketitle

\section{Introduction}

In recent years, YouTube has gained huge popularity among content creators. A large number of content creators upload their video content on this platform. These videos get tons of views and comments. The content creators, more generally called YouTubers, need to continuously work on maintaining the quality and quantity of their contents. To do so, they must collect feedback from their viewers through the comments section. This feedback lets them understand the influence of their creations. In addition to improving audience engagement, feedback also provides information on the aspects of the content that need improvement. 

However, not all YouTubers have enough time to go through all the comments on individual video. On the contrary, they must read all the comments to fully understand the public interest on their content. The solution to this inconvenience is addressed in our work. Our approach is to extract all the comments from a video and categorize them into multiple categories based on both sentiment and sentence types: Negative, Positive, Interrogative, Imperative, Corrective, and Miscellaneous. These categories can help YouTubers focus only on those comments that suit their interest.

There have been multiple studies in the field of sentiment analysis such as Twitter sentiment analysis \cite{agarwal2011sentiment}, YouTube polarity trend analysis \citep{krishna2013polarity}, user comment sentiment analysis on YouTube \citep{bhuiyan2017retrieving}, and so on. However, not enough research has been carried out on sentiment analysis through classification of a sentence based on its type. We have approached this issue from the perspective of YouTube comments. Consequently, it is a challenging task to categorize the comments into different sentence types because of various factors such as non-standard language, spelling errors, unformatted texts, and trivial comments. Apart from these, sometimes there are multiple sentences of different classes on a single comment. The combination of these issues poses a unique challenge in sentiment analysis based on sentence types.

One of the simplest ways to address the problem is to categorize the comments purely based on lexicon \citep{aung2017sentiment} e.g., the interrogative comments can be identified from keywords such as what, how, and why. Similarly, positive sentences can be identified from keywords like good, best, and wonderful. However, this approach is naive and does not address unique challenges presented by informal texts. Moreover, this method performs poorly if a single comment comprises of multiple categories. Such comments can be be categorized more efficiently by appropriately extracting features from the text corpus and using supervised machine learning techniques \citep{kadhim2019survey}. Neural networks \citep{zhou2015c} can be used as a potential solution, however, they are difficult to tune and are not readily explainable. Explainability is especially important for comments that fall under multiple categories to clearly understand why a resulting category was selected.

Our approach is to extract features from preprocessed data and use those features to train well-known supervised learning algorithms. 
The supervised learning models can then be fine tuned to get the best results. Since the performance of the model is dependent on the text corpus, we select multiple popular fine tuned algorithms for this task and observe their performance.
We experiment our YouTube comments dataset with five different fined tuned classification models using two different feature extraction methods to obtain the best results for each model. The accuracy of the models are calculated in terms of cross validation score and $F_1$ score. 
Although our approach is simple, the results are effective enabling content creators to view their feedback of interest easily.


\begin{table*}[ht]
    \caption{Summary of the related work and our approach}
    \label{tab:comparison}
    \begin{tabular}{l l l l}
    \toprule
    Study & Dataset & Categories & Methodology \\
    \midrule
    Poche~\emph{et al.}~\cite{poche2017analyzing} & Comments on coding tutorial videos & Content Concerns, Miscellaneous & SVM, Naive Bayes \\
    Taboada~\emph{et al.}~\cite{taboada2011lexicon} & Epinions 1, Epinions 2, & Setiments & Lexicon \\
                                                    & Movie reviews, Camera reviews &  & \\
    Siersdorfer~\emph{et al.}~\cite{siersdorfer2010how} & YouTube comments & Sentiments & Linear SVM + thesaurus \\
    Krishna~\emph{et al.}~\cite{krishna2013polarity} & IMDB dataset & Sentiments & Naïve Bayes \\
    Bhuiyan~\emph{et al.}~\cite{bhuiyan2017retrieving} & YouTube Comments & Sentiments & Lexicon \\
    Maas~\emph{et al.}~\cite{maas2011learning} & IMDB reviews & Sentiment + Semantic & Unsupervised + Supervised models \\
    Madden~\emph{et al.}~\cite{madden2013classification} & YouTube comments & Numerous & Manual \\
    Cheung~\emph{et al.}~\cite{cheung2005feature} & Webclopedia & Question types & Naïve Bayes, Decision Tree \\
    Kim~\cite{kim2014convolutional} & Multiple & Sentiments & Convolutional NN \\
    Hassan~\emph{et al.}~\cite{hassan2017deep} & IMDB reviews,  & Sentiments & Recurrent NN \\
    & Stanford Sentiment Treebank & & \\
    Khoo~\emph{et al.}~\cite{khoo2006experiments} & Email conversation & 14 different sentence types & Naive Bayes, Decision Tree, SVM\\
    \midrule
    Our approach & YoutTube comments & Sentence type + Sentiments & Linear SVC, Logistic Regression, \\
    &&& Multinomial NB, Random Forest, \\
    &&&  Decision Tree \\
    \bottomrule
\end{tabular}
\end{table*}

\subsection{Organization}

The rest of the paper is organized as follows: Section 2 reviews the related work on feature extraction from text, YouTube comment sentiment analysis, and sentence type categorization. Section 3 discusses our approach. Similarly, Section 4 presents the results of the approach and Section 5 concludes the paper with some remarks and future enhancements.

\section{Related Work}
Various studies have been done on feature extraction from text. In their thesis work, Poche~\emph{et al.}~\cite{poche2017analyzing} have studied the commenting behaviour of viewers on coding tutorial videos. Using SVM and Naive Bayes models, the comments collected were classified into two broad categories: Content Concerns and Miscellaneous. 
However, classifying comments into multiple sentence types is a more complex task; a detailed classification of "Content Concerns" was performed in our work with increased accuracy.

Taboada~\emph{et al.}~\cite{taboada2011lexicon} implemented lexicon based methods for sentiment analysis. In this study, a dictionary of words was created and the words were manually ranked into a scale of -5 to +5 (negative to positive sentiment). 
Our supervised learning approach replaces the manual work performed during the classification step of their approach, resulting in a greater accuracy. The time consuming process of manually scaling the words was also automated by our machine learning approach. Our research classifies texts based on the type of sentence as well as the sentiment. 
Siersdorfer~\emph{et al.}~\cite{siersdorfer2010how} used linear support vector machine along with a thesaurus to obtain the degree of negativity and positivity of each word from comments. The accuracy of their approach stands at 0.72. Even without using any thesauruses, we were able to increase the accuracy of our model for the sentiment analysis.

Likewise, Krishna~\emph{et al.}~\cite{krishna2013polarity} used Naïve Bayes classification for polarity (sentiment) analysis on IMDB dataset. 
Their approach makes use of the category of each comment during the feature selection process, which means it cannot calculate the polarity of a new comment because its features cannot be calculated without knowing its category. However, our approach does not use category during feature selection process. In a slightly different approach, the sentiment on comments is used to aid in finding the relevant video on YouTube based on the search text by Bhuiyan~\emph{et al.}~\cite{bhuiyan2017retrieving}. 
This approach was only able to classify the comments into positive and negative categories.

Maas~\emph{et al.}~\cite{maas2011learning} introduced an advanced concept by showing the relationship among the words in their logical meaning as well as sentimental sense. The semantic aspect was trained by an unsupervised model, whereas the sentiment aspect was trained by a supervised model (SVM). Although their model gave the polarity of the text, they did not experiment with the type of sentence. Using the conventional NLP tools, we were able to classify text based on both sentiment and type of sentence.

When it comes to classifying by sentence types, Madden~\emph{et al.}~\cite{madden2013classification} provided a scheme for classification of YouTube comments. They proposed that people’s commenting habits differ between groups; some comment for promotion, some for providing information, and others for pleasure. Madden's work has provided 10 main categories and 58 sub categories for such comments. However, they have merely provided a schema for the classification task. We attempted to use some of their categories into the classification task. Similarly, Cheung~\emph{et al.}~\cite{cheung2005feature} classified questions for the data obtained from Webclopedia. 
Their work was limited to the classification of different types of question.

Kim~\cite{kim2014convolutional} implemented convolution neural networks for analysing sentiment on 7 different datasets. The paper showed that hypertuned Convolution Neural Network (CNN) can produce satisfying results. The word vectors were obtained from an unsupervised neural language model~\emph{Word2vec} \citep{collobert2011natural, socher2011semi, iyyer2014political}.
The variants of CNN models used were CNN-rand, CNN-static, CNN-non-static and CNN-multichannel. 
On the other hand, Hassan~\emph{et al.}~\cite{hassan2017deep} proposed a recurrent neural network based model called Long Short-Term Memory (LSTM) for sentiment analysis. 
Both of these approaches showed promising results, but they have not explored the classification of different types of sentences (like imperative, question, corrective, and sentimental). They are likewise limited to either sentiment analysis or question classification. Our research addresses much broader sentence classification.

In most of the existing work, the sentences of the imperative class have not been researched adequately. Khoo~\emph{et al.}~\cite{khoo2006experiments} performed experiments on different models for 14 different classes of sentences (including imperative sentence types like request, instruction and suggestions). The models used for the experiment were Naive Bayes, Decision Tree, and Support Vector Machine. 
Support Vector Machine overshadowed all other models and had insignificant effect from feature selection. In their work, they only chose the standard response emails because these emails have well-structured sentences and few grammatical errors. It eases the classification task. However, our work consists of a large number of unstructured sentences with huge grammatical errors. Table.~\ref{tab:comparison} shows the comparative study of the previous works and our approach.

\section{Methodology}
In our approach, we make use of python programming language as in \citep{ahuja2019} to execute the experiment. We collect comments from YouTube videos using the YouTube Data API, which are preprocessed to filter out the noise. Then, the features are extracted from the comments and fed into the classification models. The models are finally hypertuned to produce the best results.

\subsection{Data Collection}
\begin{figure}[b]
    \vspace{-20pt}
    \centering
    \includegraphics[width=0.4\textwidth]{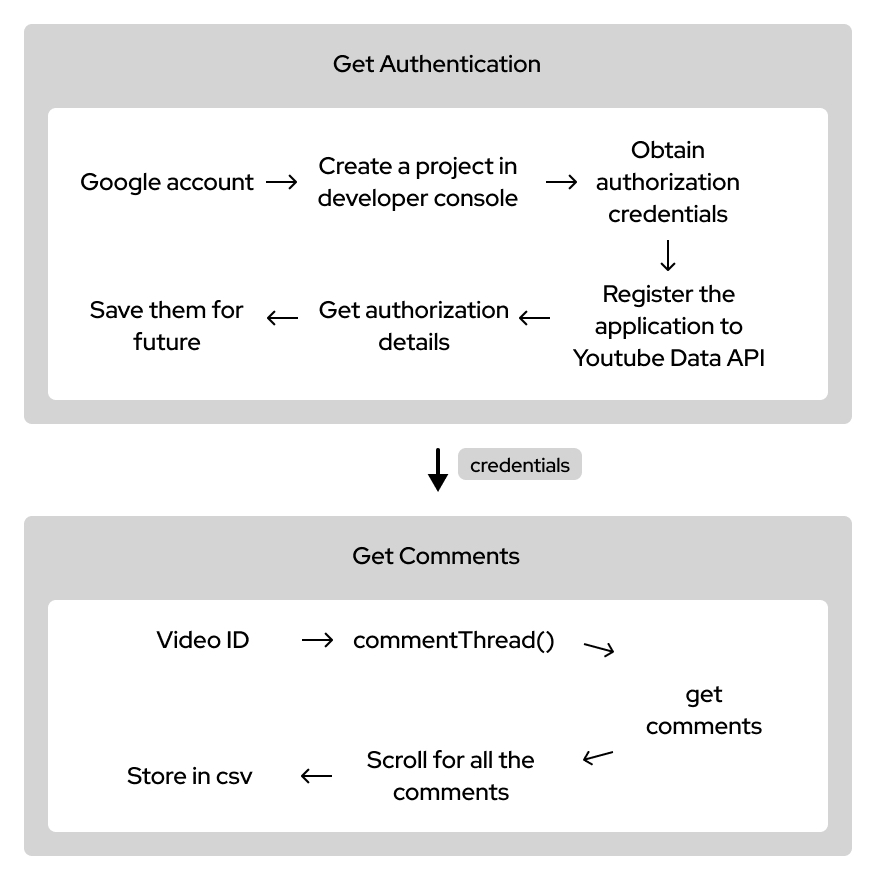}
    \caption{The process of scraping}
    \label{fig:scraping}
\end{figure}

We build our dataset by scraping YouTube comments. We use the YouTube Data API for authenticating and accessing the comments on videos \citep{yasima2016using}. First, the API is used for authentication and then the credentials obtained are fed into the comment extractor. The comment extractor then scrapes comments from the comment section by scrolling through all the comments and loading them dynamically. Fig.~\ref{fig:scraping} shows the scraping process.

The dataset consists of 10,000 comments picked from different tutorial videos~\footnote{Source Code and Dataset Publicly Available at \url{https://github.com/Rhitabrat/Youtube-Comments-Categorization}}. We chose tutorial videos for our experiments because they contain a wide variety of comments. We then manually label the comments into 6 different classes: Positive, Negative, Interrogative, Imperative, Corrective and Miscellaneous. These classes are defined based on general needs of the YouTubers. Note that further categories can be established if or as needed. These classes belong to two broader classes: Sentiment Analysis (positive and negative) and Sentence Types (Interrogative, Imperative, Corrective and Miscellaneous). Table.~\ref{tab:classes} shows different classes and the content of that class. The classes are explained in more detail next.

\begin{table}[t]
    \caption{Classes of comments with content type}
    \label{tab:classes}
    \begin{tabular}{c l}
    \toprule
    Class&Content\\
    \midrule
    Positive        & appraisals, appreciations \\
    Negative        & scoldings, not able to do what is told in the \\
                    & video \\
    Interrogative   & all type of questions, queries, asking for \\
                    & something, sentences starting with \\
                    & modal/auxiliary verbs \\
    Imperative      & requests, commands, expectations \\
    Corrective      & amendments, improvements, mistakes, \\
                    & remedy \\
    Miscellaneous   & remaining types, promotions, chitchat \\
    \bottomrule
\end{tabular}
\end{table}

\emph{Positive} tells that the viewers perceived the content as worthy and that the content created a positive impact on them. \emph{Negative} provides information on what is wrong with the content and why the viewers are not attracted to it. \emph{Interrogative} conveys viewers' doubts and questions. It is a useful feature because the content creators can increase their influence by addressing viewers’ questions and issues. \emph{Imperative} provides viewers’ expectations and requests for actions. \emph{Corrective} tells the creator about the corrections expected to be made on the video making it a very important aspect of the feedback. \emph{Miscellaneous} includes declarative sentences and all other trivial comments. We manually labelled the miscellaneous category from the creators' point of view.

\begin{figure}
    \centering
    \includegraphics[width=0.45\textwidth]{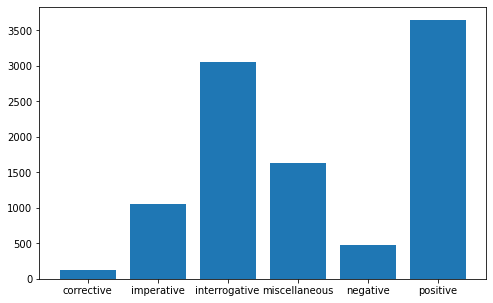}
    \caption{Number of comments in each class}
    \label{fig:barplot}
\end{figure}

As mentioned previously, some of the comments can belong to more than one class. For instance, the first sentence in "Your solution is not practical. Can you suggest another one?", suggests the negative class while the second sentence suggests the interrogative class. In such situations, we classified the comment based on the importance to the content creator. In this example, we assumed it as an interrogative sentence because it is more important to answer the question to increase odds of the viewer to return and stay engaged in the content of the channel. Fig.~\ref{fig:barplot} shows the visual presentation of the quantity of comments in each class.

\subsection{Data Pre-processing}

It is important to clean the data and have them in appropriate format to improve classification. The data pre-processing step handles the following factors that make the classification process difficult:
\begin{enumerate}
    \item Non-standard language: The texts used in the comments section do not always employ standard English. Comments often contain slangs and improper form of words, making it difficult to extract features from them.
    \item Spelling errors:  Commenters often do not pay attention to spellings due to the informal setting. Such spelling errors need to be corrected; otherwise, the words would add unnecessary features to the classification model, decreasing the overall accuracy of the classifier and impacting efficiency of the approach. For example, the words "plz" and "please" convey the same meaning in informal writing, but if the incorrect spelling is ignored, they would be treated as two different words by the classifier.
    \item Unformatted texts: These refer to comments containing computer codes. These do not contribute to the feature extraction accuracy; rather, they add unnecessary load to the feature matrix.
    \item Trivial comments: Not all the comments posted were about the video or related to the channel. A large number of viewers comment in order to market their products or just to show their presence. These comments are not useful to the content creators and only add unnecessary overhead.
\end{enumerate}

Above issues are common in platforms like YouTube because of the informal nature of communication. We addressed these issues using the following pre-processing steps:
\begin{itemize}
    \item lowercasing
    \item removing URLs
    \item removing new line character ("$\backslash$n")
    \item removing punctuations
    \item removing integers
    \item removing emojis
    \item correcting spelling errors
    \item lemmatizing
    \item removing stopwords
\end{itemize}

Given the nature of this study, lowercasing was relevant because the same word would have been identified as a different feature had some letters been capitalized (for example, "Love" and "love" are two different words for a computer). The removal of URLs, new line characters, punctuation, integers, and emojis was performed because they did not provide useful information for feature extraction; rather adding unnecessary complexity to the model.

Since there are a lot of spelling errors in informal writing, we corrected the typos using the autocorrect-library from Python. We used Lemmatization to analyze the words morphologically and group the similar words together. Furthermore, there are certain frequent words that do not add significant meaning to the sentence such as "is", "are", and "it". They were removed from the corpus. However, stopwords were not removed from all the classes of comments because stopwords for one category might be important for another category. For example, "not" and "no" are important for the negative class whereas they are not important for other classes. Stopwords were used from nltk English corpus, which consists of 179 stopwords. Table.~\ref{tab:stopwords} shows the stopwords that were not considered in each category.

\begin{table}[b]
    \caption{Stopwords ignored in each class}
    \label{tab:stopwords}
    \begin{tabular}{c l}
    \toprule
    Class&Ignored Stopwords\\
    \midrule
    Positive        & NA \\
    Negative        & no, not \\
    Interrogative   & how, what, which, who, whom, why, do, is, \\
                    & does, are, was, were, will, am, are, could, \\
                    & would, should, can, did, does, do, had, have \\
    Imperative      & could, would, should, can \\
    Corrective      & NA \\
    Miscellaneous   & NA \\
    \bottomrule
\end{tabular}
\end{table}

\subsection{Converting text into features}
Once pre-processing is completed, the pre-processed text is filtered so that only the necessary features remain in the corpus. This helps in reducing the load to the classifiers and also increases the accuracy of selected models \citep{yang1997proceedings}. 

The well-known techniques for vectorizing a corpus of text include document frequency, tf-idf vectorizer, hashing vectorizer, and Word2Vec. We selected document frequency vectorizer and tf-idf vectorizer for this paper. Using these two methods we can study the behaviour of different classification models under two different conditions. Document frequency (\emph{df}) vectorizer gives importance to the term that has higher frequency in the document; whereas, \emph{tf-idf} can incorporate the terms that are rarely present in the document. Unlike hashing vectorizer, we can examine the text features which are important to the model using the vectors generated by~\emph{df} and~\emph{tf-idf}. For rare or out of vocabulary terms (which might be important to a model),~\emph{Word2Vec} can not create an ideal vector for them and it is difficult to interpret those vectors because of hidden layers.


When calculating document frequency~(Eqn.~\ref{eqn:df}), if the same term is present multiple times in a comment, then its additional counts are not considered. Also, the terms that appear in less than or equal to 5 comments are ignored because they do not add value to the features. In the same way, if any term appears in the majority of the comments, it does not add value to the feature because it is not the distinguishable feature for a class. These terms are likely already filtered by the stopwords removal process. However, we ensure that only terms that significantly add value to their comment’s class are considered.

\begin{equation}\label{eqn:df}
    df = \frac{n}{N}
\end{equation}
where, $\ df $ denotes document frequency, $\ n $ denotes number of documents in which the term appears, and $\ N $ denotes total number of documents where document means comment. 

After above steps, 2210 terms (features) were derived and scaled from 0 to 1 using a min-max scalar (normalization). We performed this because some of the machine learning models cannot handle large ranges of data. Doing so also helps in speeding up some of the calculations.

\begin{equation}
    Min Max Scaler(x) = \frac{x - x_{min}}{x_{max} - x_{min}}
\end{equation}
where, $\ x $ is observed value, $\ x_{min} $ is the minimum value of that class and $\ x_{max} $ is the maximum value of that class.
   
The second feature extraction technique used in this paper is~\emph{tf-idf} (term frequency - inverse document frequency). It not only considers the frequent terms, but also the rare terms.

\begin{equation}
    tf{\text -}idf = tf * \log \frac{1}{df}
\end{equation}
where, $\ tf $ is the term frequency and $\ df $ is the document frequency

For~\emph{td-idf}, we got 4304 features when both unigram and bigram were taken into account.

\begin{figure}[b]
    \centering
    \includegraphics[width=0.2\textwidth]{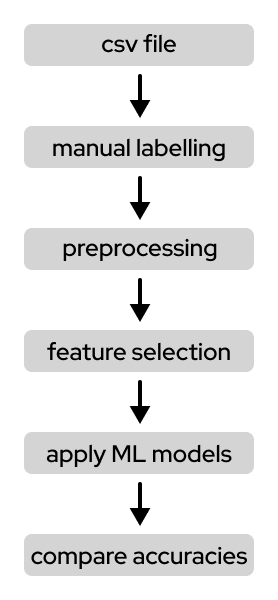}
    \caption{The complete classification process}
    \label{fig:process}
\end{figure}

\subsection{Model Selection and Hyperparameter Tuning}
Each machine learning model has its own strengths and weaknesses when dealing with a text corpus. The fitness of a model for a specific dataset depends on the characteristics of the model as well as the features of the dataset. Since our text corpus is dense and has numerous features which significantly affects the classification task, the machine learning models selected are based on the density of the features and the number of classification categories (binary or multiple). Linear Support Vector Classification (Linear SVC), Logistic Regression, Multinomial Naive Bayes (Multinomial NB), Random Forest Classifier, and Decision Tree Classifier are selected as we work with a dense dataset and multiple classes.  Especially, Random Forest Classifier helps prevent the overfitting problem and mitigates the impact of outliers, whereas Decision Tree Classifier can easily deal with the irrelevant features on the dataset.

For each model selected, the outcome can be enhanced by optimizing the hyperparameters. To understand how different parameters can significantly affect the performance of each model, we experimented all the models with the minimum subsets of those parameters (grid search strategy). Table.~\ref{tab:hyperparameter_tuning_tfidf} shows all the values of the parameters that were tested for both~$df$ and~$tf-idf$ measures. The columns "Best Value for" present the value with the best result, which were used for the final experiment. 

\begin{table*}
    \caption{Parameters selected for tuning $df$ and $tf{\text -}idf$ features}
    \label{tab:hyperparameter_tuning_tfidf}
    \centering
    \begin{tabular}{c c l c c}
    \toprule
    Model & Parameters & Values Experimented & Best Value for~\emph{df} & Best Value for~\emph{tf-df}\\
    \midrule
    Linear SVC          & C         & 50, 10, 1.0, 0.1, 0.01        & 1.0 & 1.0 \\
    Logistic            & C         & 100, 10, 1.0, 0.1, 0.01       & 1.0 & 10 \\
    Regression          & solvers   & newton-cg, lbfgs, liblinear   & newton-cg & liblinear \\
                        & penalty   & l1, l2, elasticnet, none      & l2 & l2 \\
    Multinomial NB      & fit\_prior & True, False                  & True & False \\
                        & alpha     & 0,0.5,1                       & 1 & 1 \\
    Random Forest Classifier      & n\_estimators & 10, 100, 1000             & 1000 & 1000 \\
                        & max\_features      & sqrt, log2           & log2 & log2 \\
                        & criterion         & gini, entropy         & entropy & entropy \\
    Decision Tree Classifier      & criterion         & gini, entropy         & gini & entropy \\
                        & max\_features      & sqrt, log2, auto, None & None & None  \\
    \bottomrule
\end{tabular}
\end{table*}

\section{Results}
\subsection{Experimental Setup}
All experiments were written in Python programming language and executed using Jupyter Notebooks. The ``pandas" library was used for importing the data. The pre-processing step was carried out by regex functionality from the ``re" library. Stopwords were removed by the ``nltk" library. Similarly, spell correctors used the ``autocorrect" library from Python.  The feature selection and classification processes used the ``sklearnmodule".  For the training purpose, 80\% of comments were randomly selected from the dataset, and the remaining 20\% were used for testing. The experiments were run on Google Colab with 16.29 GB of RAM and 2.20GHz Intel(R) Xeon(R) CPU with GPU turned off.
The overall process is shown in Fig.~\ref{fig:process}.

\subsection{Result Analysis}
We study the accuracy of results obtained from all the five models with two different feature extraction techniques in this section. The cross validation score ranged from 0.76 to 0.85, whereas $F_1$ score ranged from 0.81 to 0.87. Let us analyze the performance of the classification models. For instance, we chose  the results of the Random Forest classifier with document frequency as a feature selector. Some of the correctly classified comments for Random Forest classifier are shown in Table.~\ref{tab:correct_comments} and the incorrectly classified samples are shown in Table.~\ref{tab:incorrect_comments}.

\begin{table}[b]
    \caption{Correctly classified comments}
    \label{tab:correct_comments}
    \begin{tabular}{c l c}
    \toprule
    Index & Comment & Predicted Class \\
    \midrule
    200     & hi, how to do this environment    & interrogative \\
            & setting on windows 10             &   \\
    201     & Thank you sooo much , sir!        & positive \\
    202     & God has manifested to me in your  &                      \\
            & form. Thank you.                  & positive     \\
    203     & thank you sir helped alot.....    & positive   \\
\end{tabular}
\end{table}

\begin{table}
    \caption{Incorrectly classified comments}
    \label{tab:incorrect_comments}
    \begin{tabular}{c l c}
    \toprule
    Index & Comment & Predicted Class \\
    \midrule
    133     & Sir please teach these concepts    & positive      \\
            & using code...I really request you &   (imperative)               \\
            & to explain it with code..Thank    &                  \\
            & you so much for teaching all      &     \\
            & these concepts so we...           &                \\
    135     & Time complexity for pure     & miscellaneous \\
            & recursion should be pow(2, m+n).  & (corrective)              \\
            & Correct me if I am wrong          &              \\
    213     & You are an amazing Teacher...       & miscellaneous \\
    && (positive)\\
    315     & Your repeated the same meaning      & miscellaneous \\
            & for about 1 minute                 &               (negative)               \\
\end{tabular}
\end{table}

In Table.~\ref{tab:correct_comments}, the first comment had a distinguishing term "how" that implies interrogative class. Likewise, the second comment consisted of the term "Thank you" which suggested positive sentiment. Thus, these were correctly classified by the model.

The first two incorrectly classified comments in Table.~\ref{tab:incorrect_comments} showed that the classifier gets confused when there are two types of sentences in the single comment. In the first comment, the first part of the sentence suggested imperative class, whereas the second part suggested a positive class. Earlier, when labelling the data, we classified the comment into imperative class because the content creator prioritized an imperative class over the positive class as they needed to focus on the grievances of the viewers (which was given by imperative class). Similarly, the third comment in Table.~\ref{tab:incorrect_comments} was falsely classified. The most important word in this comment for distinguishing the category was "amazing". This word suggests the positive sentiment. Unfortunately, the classifier could not detect it because the corpus contained only a few positive sentences that included this term. 

The results of the best five models are shown in Table.~\ref{tab:accuracy_df} and Table.~\ref{tab:accuracy_td_idf}. We can see that all the five models performed satisfactorily except that Decision Tree Classifier showed weaker performance for both the feature selection methods. This is because of a high dimensional data (i.e. having dense features). This classifier splits the feature space at each node; as the feature space increases, the distance between the feature points also increase which makes it difficult to find a good split. Random Forest also works on the similar concept of Decision Tree, but it showed good performance because it does not use all the features rather it uses a collection of decision trees. This helps to reduce the distance between the the data points. 

\begin{table}[b]
    \caption{\emph{df} used as a feature extraction method}
    \label{tab:accuracy_df}
    \begin{tabular}{c c c}
    \toprule
    Model Name & Cross Validation Score & F1-Score \\
    \midrule
    Linear SVC                  & 0.83      & 0.86  \\
    Logistic Regression         & 0.85      & 0.87  \\
    Multinomial NB              & 0.79      & 0.84  \\
    Random Forest Classifier    & 0.83      & 0.85  \\
    Decision Tree Classifier    & 0.80      & 0.83  \\
\end{tabular}
\end{table}

\begin{table}[t]
    \caption{\emph{tf-idf} used as a feature extraction method}
    \label{tab:accuracy_td_idf}
    \begin{tabular}{c c c}
    \toprule
    Model Name & Cross Validation Score & F1-Score \\
    \midrule
    Linear SVC                  & 0.84      & 0.86  \\
    Logistic Regression         & 0.84      & 0.86  \\
    Multinomial NB              & 0.76      & 0.82  \\
    Random Forest Classifier    & 0.83      & 0.84  \\
    Decision Tree Classifier    & 0.80      & 0.81  \\
\end{tabular}
\end{table}

\begin{figure}[b]
    \centering
    \includegraphics[width=0.45\textwidth]{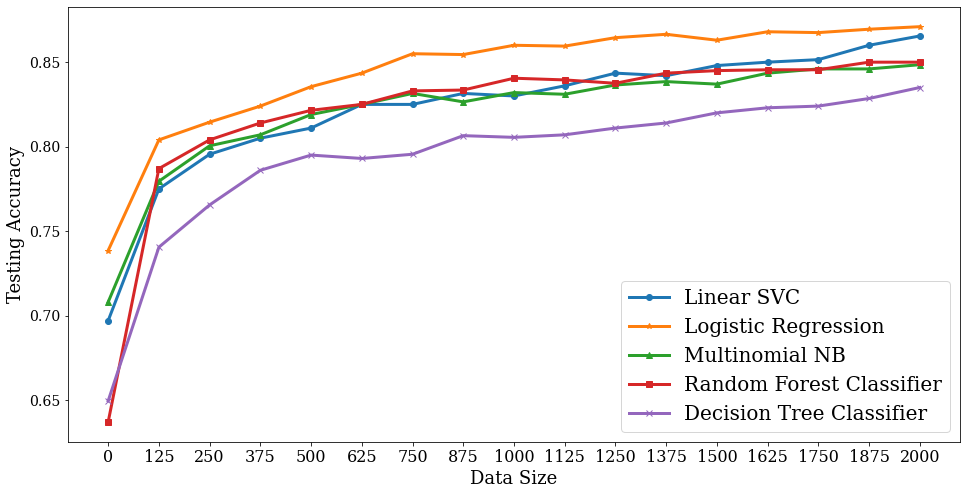}
    \caption{Testing Accuracy vs Data Size plot for five different models ($df$)}
    \Description{The accuracy of each model is increasing with the increasing in the data, when document frequency is used as feature extraction method.}
    \label{fig:accuracy_df}
\end{figure}

\begin{figure}[b]
    \centering
    \includegraphics[width=0.45\textwidth]{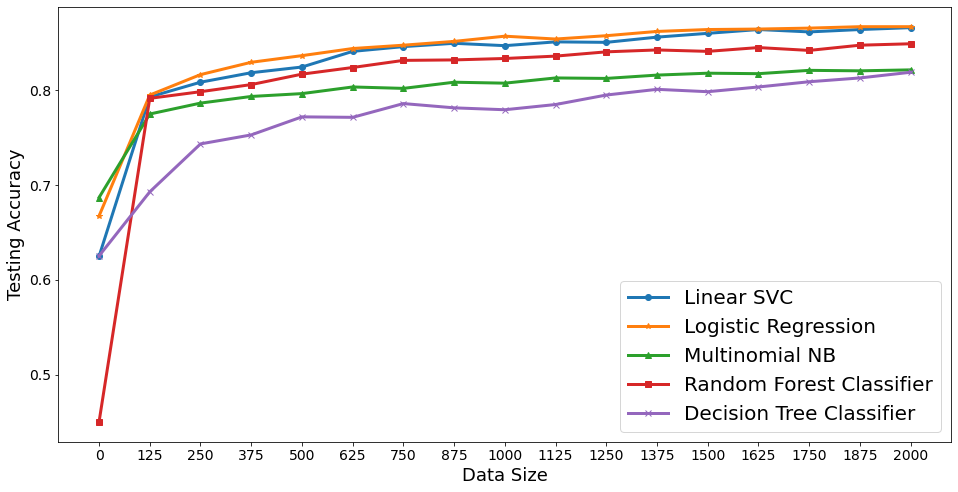}
    \caption{Testing Accuracy vs Data Size plot for five different models ($tf-idf$)}
    \Description{The accuracy of each model is increasing with the increasing in the data, when tf-idf is used as feature extraction method.}
    \label{fig:accuracy_tf_df}
\end{figure}

The reasonable cross validation scores for the models show that the model is not overfitted with the training data. Similarly, the spelling corrections performed on the training data had very insignificant impact on the performance of the models. The impact was only around 0.01\%. Additionally, Fig.~\ref{fig:accuracy_df} and Fig.~\ref{fig:accuracy_tf_df} show that the accuracy of each model increased with the increase in the data size.

\begin{figure*}
    \centering
    \includegraphics[width=0.9\textwidth]{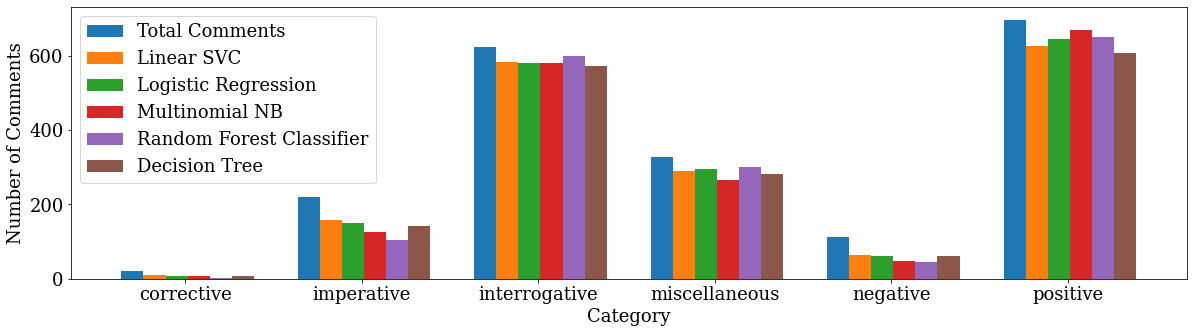}
    \caption{Performance of all the models in individual category for document frequency}
    \label{fig:accuracy_comparison_df}
\end{figure*}
\begin{figure*}
    \centering
    \includegraphics[width=0.9\textwidth]{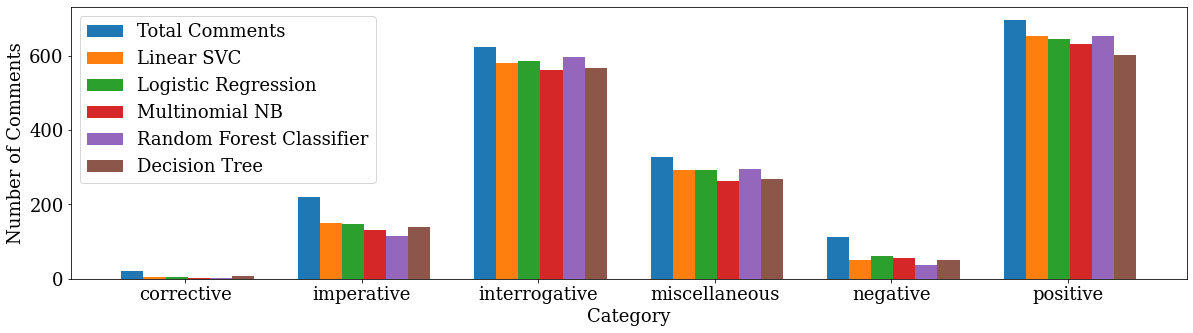}
    \caption{Performance of all the models in individual category for TF-IDF}
    \label{fig:accuracy_comparison_tf_idf}
\end{figure*}

As shown in Fig.~\ref{fig:accuracy_comparison_df} and Fig.~\ref{fig:accuracy_comparison_tf_idf}, 
the poor performance in miscellaneous category can be inferred to the presence of diverse types of sentences that are classified under miscellaneous category, which confuse the models. The models perform relatively poor in imperative and negative categories as well. The imperative comments consisted of certain terms that fall under interrogative category, which complicated the classification task. For instance, "Would you please explain about library functions?" consisted of the terms "please", an imperative term; and "Would", an interrogative term. Likewise, the negative comments could contain both the interrogative part as well as the negative part. For example, "You are useless. Why do you keep repeating the same thing?". These types of comments create ambiguity, which result in degraded performance in the aforementioned categories. 

\section{Conclusion}

With the successful classification of the comments into their respective categories, a YouTuber can easily access each category of comment. The positive and negative category show the public sentiment on the video. Other categories aid the YouTuber to distinctly view the questions asked about the video and the suggestions provided to improve the content. This can help the YouTuber to avoid scrolling through hundreds of comments and filtering them manually for each video. 
While previous researchers focused either on sentiment analysis or classification of sentence of a niche, we have incorporated both the aspects. In this paper, we classified the comments using 5 different models on 2 feature selection methods. The experiments showed that best scores for cross validation and $F_1$ were obtained by Logistic Regression.

In future work, the number of classes and sub-classes can be increased to represent a more comprehensive comment classification. Likewise, the classification models and overall feature selection approach can be further improved for the comments that belong to more than one class. 
We also plan to extend this research by performing a comparative study with Explainable Neural Networks (xNNs) for comments classification.

\bibliographystyle{ACM-Reference-Format}
\bibliography{sample-base}

\end{document}